\documentclass[12pt]{article}
\usepackage{epsf}

\newcommand{\eeq}{\end{equation}}

\newcommand{\beq}{\begin{equation}}
\newcommand{\beql}{\begin{eqnarray}}
\newcommand{\eeql}{\end{eqnarray}}

\begin{document}

\title{ Counting Multiple Solutions in 
Glassy Random Matrix Models }
\author{ 
N. Deo$^{1,2}$,\\
$^1$ Poornaprajna Institute for Scientific Research,\\ 
Bangalore 560080, India,\\
$^2$ Abdus Salam International Centre for Theoretical\\
Physics, Trieste, Italy.\\}

\maketitle

\begin{abstract}

This is a first step in counting the number of multiple solutions in 
certain glassy random 
matrix models introduced in refs. \cite{d02}. We are able to do this by 
reducing 
the problem of counting the multiple solutions to that of a moment problem.
More precisely we count the number of different moments when
we introduce an asymmetry (tapping) in the random matrix model and 
then take it to vanish. It is shown here that the number of moments 
grows exponentially with respect to N the size of the matrix.
As these models
map onto models of structural glasses in the high temperature phase (liquid)
this may have interesting implications for the supercooled liquid phase in
these spin glass models. Further it is shown that the nature of the asymmetry
(tapping) is crutial in finding the multiple solutions. This also clarifies 
some of the puzzles we raised in ref. \cite{bd99}.   

\end{abstract}

PACS: 02.70.Ns, 61.20.Lc, 61.43.Fs


\section{Introduction}\label{intro}

Random matrix models can be used very effectively as simple mathematical toy 
models where many new ideas in physics, biology and economics can be tested 
analytically ref. \cite{mg91,oz01,bs99}. 
Here we try to understand the idea of tapping and 
counting, well studied in the context of granular media, in the glassy random 
matrix model introduced in ref. \cite{d02}. There it was demonstrated
that the matrix models with gaps in their eigenvalue distribution had multiple
solutions and were related to the high temperature phase of certain p-spin 
glass models ref. \cite{ckpr95}. We approach the problem in much the same 
spirit as done for spin systems in ref. \cite{dl01}. This is a first step in 
understanding what happens when we tap the model ie introduce a perturbation 
and remove it. This enables us to count the number of different configurations.
Studies to understand the fluctuation-dissipation
relations and the relations between the dynamical and Edwards temperature 
in the dynamical matrix models awaits further work. This study will also 
help us understand some of the puzzles that we raised in ref. \cite{bd99}. 
One of the puzzles in these models is that the long range correlators 
found in ref. \cite{aa96} by mean field calculations differ from that
found in ref. \cite{d97,bd99} using the orthogonal polynomial methods.
A resolution of this has been suggested in ref. \cite{bde00} where it is
claimed that the difference arises due to discreteness of the number of 
eigenvalues for double well models with equal depths. Here we try to 
understand these results using the method of moments.    

Most of the studies and applications of matrix models correspond to eigenvalue
distributions on a single-cut in the complex plane where the eigenvalue 
density is non-zero ref. \cite{mg91}. Here we study a one hermitian matrix 
model 
with a more complicated eigenvalue structure. These have found applications in 
two-dimensional quantum gravity, string theory, disordered condensed matter
systems, superconductors (with complex vector potential and with impurities)
and glasses. Here we study these models with applications to glasses in mind
as discussed in refs. \cite{d02}. To illustrate some of the generic 
properties we
study a one hermitian matrix model with two cuts for the eigenvalue density.
One of the important differences observed in these models is that they have
multiple solutions which show up in certain correlation functions. Here we
count the number of multiple solutions and explore the possibility that these
multiple solutions arise by taking different paths in phase space ( each path
may correspond to a different metastable glassy state ). It is important to
establish the correspondence between the multiple solutions and metastable
glassy states. The barrier heights corresponding to these various solutions
are also future goals.

I will discuss here the matrix model with double-well potential the $M^4$
model (in the Gaussian Penner model where similar things happen will be 
pursued elsewhere). A tapping is introduced which corresponds to coupling the
matrix model to an external source. The limit of taking the external
sources to vanish gives different values for the moments in these models.
This may result in different values for the partition function and hence the
free energy. Taking different tappings corresponds to exploring the full
space of configurations. Here we present the first steps in counting 
the number of different
configurations and find it to be exponentially large.

After this work was completed
we find that in a different context results of exponentially 
large number of minima have been reported in a renormalizable 
matrix potential with $S_N$ using a different method by Soljacic
and Wilczek ref. \cite{sw00}.

\section{Notations and Conventions}\label{note}

Let $M$ be a hermitian matrix. The partition function to be
considered is $ Z=\int dM e^{-N tr V(M)} $
where $M=N\times N$ hermitian matrix. The Haar measure 
$dM = \prod_{i=1}^{N}dM_{ii}\prod_{i<j} dM_{ij}^{(1)}
dM_{ij}^{(2)}$ with $M_{ij}=M_{ij}^{(1)} + i M_{ij}^{(2)}$ 
and $N^2$ independent variables. $V(M)$ is a polynomial in M:
$
V(M)=g_1 M + (g_2/2) M^2 + (g_3/3) M^3 + (g_4/4) M^4 + .....
$
The partition function is invariant under the change of variable
$M^{\prime}=U M U^{\dagger}$ where $U$ is a unitary matrix. We can 
use this invariance and go to the diagonal basis ie 
$D^{\prime}=U M U^{\dagger}$ such that $D^{\prime}$ is the matrix
diagonal to $M$ with eigenvalues $\lambda_1,\lambda_2,.....\lambda_N$.
Then the partition function becomes
$
Z = C \int_{-\infty}^{\infty} \prod_{i=1}^{N} d\lambda_i 
\Delta (\lambda)^2 e^{-N \sum_{i=1}^N V(\lambda_i)}
$
where $\Delta (\lambda) = \prod_{i<j} |\lambda_i-\lambda_j|$
is the Vandermonde determinant. The integration over the group
U with the appropriate measure is trivial and is just the constant C. 
By exponentiating the determinant as a `trace log' we arrive at the 
Dyson Gas or Coulomb Gas picture. The partition function is simply
$
Z = C \int_{-\infty}^{\infty} \prod_{i=1}^N d\lambda_i e^{-S(\lambda)} 
$
with
$ S(\lambda) = N \sum_{i=1}^N V(\lambda_i) - 2
\sum_{i,j,i\ne j} ln |\lambda_i-\lambda_j|$. 

\begin{figure} 

\leavevmode
\epsfxsize=4in
\epsffile{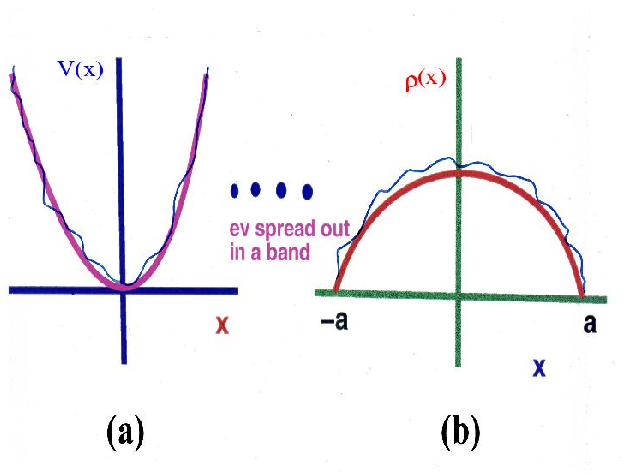}


\caption{ (a). The confining potential (b). The density of eigenvalues }
\label{fig1ab} 
\end{figure}

This is just a system of N particles with coordinates $\lambda_i$ on
the real line, confined by a potential and repelling each other with 
a logarithmic repulsion. The
spectrum or the density of eigenvalues $\rho (x) =
{1\over N} \sum_{i=1}^N \delta (x-\lambda_i)$ is in the large N limit
or doing the saddle point analysis just the Wigner semi-circle for a
(Gaussian probability distribution for the eigenvalues) quadratic
potential. The physical picture is that the eigenvalues try to be at the 
bottom of the well. But it costs energy to sit on top of each other
because of logarithmic repulsion, so they spread. $\rho$ has support on a 
finite line segment. This continues to be true whether the potential is
quadratic or a more general polynomial and only depends on there being
a single well though the shape of the Wigner semi-circle is correspondingly 
modified. For the quadratic potential the density is 
$\rho (x) = {1 \over \pi} \sqrt{(x-a)(b-x)}$ where $ [a,b] $ are 
the end of the cuts. See Fig. \ref{fig1ab}.

On changing the potential more drastically by having two humps or wells,
the simplest example being a potential 
$V (M) = - {\mu \over 2} M^2 + {g \over 4} M^4$, 
the density can get disconnected support. The precise expressions for the 
density of eigenvalues are as follows:

\beql
\rho (x) &=& {g \over \pi} x \sqrt{(x^2-a^2)(b^2-x^2)}  ~~~~~~~~   
 a<x<b \nonumber\\   
&=& 0 ~~~~~~~~ -b<x<-a 
\eeql

where $ a^2 = {1 \over g} [|\mu|-2\sqrt{g}]$ and
$ b^2 = {1 \over g} [|\mu|+2\sqrt{g}]$
and $|\mu|>2\sqrt{g}$, which is the condition that the wells 
are sufficiently deep. The eigenvalues sit in symmetric bands
centered around each well. Thus $\rho$ has support on two line
segments. As $|\mu|$ approaches $2\sqrt{g}$, $ a \rightarrow 0$ and
the two bands merge at the origin. The density is then
\beql
\rho (x) &=& {{g x^2} \over \pi} \sqrt{x^2-{2\mu \over g}}
~~~~~~~~ -\sqrt{2|\mu| \over g} < x < \sqrt{2|\mu| \over g} \nonumber \\
&=& 0 ~~~~~~~~ otherwise.
\eeql

\begin{figure} 

\leavevmode
\epsfxsize=4in
\epsffile{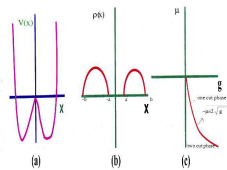}

\caption{ (a). The double-well potential (b). Density of eigenvalues 
(c). The phase diagram }
\label{fig2ab} 
\end{figure}

The phase diagram and density of eigenvalues for the $M^4$ potential 
is shown in Figs. \ref{fig2ab}.

The simplest way to determine $\rho (z)$ explicitly is to use the
generating function $F(z)=<{1 \over N} Tr {1 \over {z-M}}>$ and
its saddle point or Schwinger-Dyson equation also known in the 
mathematics literature as the Riemann-Hilbert problem 
$ F(z) = {1 \over 2} [ V^{\prime} (z) + \sqrt{\Delta} (z) ] $
with $\Delta (z) = V^{\prime} (z)^2 - 4 b(z)$
and $b(z) = g z^2 + \mu + g < {1 \over N} Tr M^2 >$ 
(see ref. \cite{bdjt93}). The density
$\rho (x)$ is then determined by the formula 
$\rho (z) = -{1 \over 2\pi} Im \sqrt{\Delta (z)}$. 
In what follows the matrix model is tapped ( that is a small perturbation
is added which breaks the $Z_2$ symmetry ) and the number of solutions
corresponding to the different moments of the model is counted. 

\section{ Introducing Asymmetry (tapping) }
\label{asym} 

Let us put a matrix source $A$, with eigenvalue $a_n$, which will
ultimately vanish in the partition function

\beq
Z_N (A) = \int dM e^{-N Tr (V(M)-AM)}.
\eeq

Using Harish-Chandra-Itzykson-Zuber 

\beq
Z_N (A) = \int \prod_1^N d\lambda_i {\Delta(\lambda) \over \Delta(a)}
e^{-\sum_1^N (V(\lambda_i)-a_i\lambda_i)}
\eeq

where

\beq
\Delta(\lambda) = Det \lambda_i^{j-1}.
\eeq

Then in terms of the moments the partition function becomes

\beq
Z_N (A) = {Det(m_n(a_k)) \over \Delta (a)}
\eeq

with 

\beq
m_n (a) = \int dx e^{-N[V(x)-ax]}x^n .
\eeq

Let us consider $m_n (a)$ if $N$ goes to infinity before $a\rightarrow 0$.

\begin{figure} 

\leavevmode
\epsfxsize=4in
\epsffile{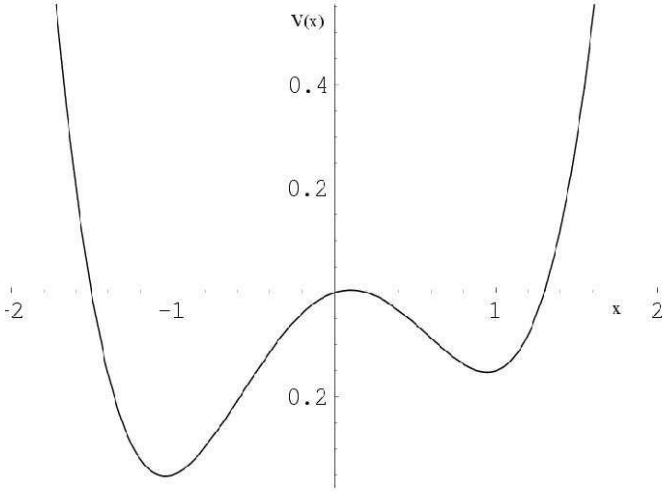}

\caption{ The Asymmetric Potential $\widetilde{V}(x)$ }

\label{fig3a} 

\end{figure}

(I). First take a non-$Z_2$ symmetric $V(x)$ with two wells see Fig. 
     \ref{fig3a}

\begin{figure} 

\leavevmode
\epsfxsize=4in
\epsffile{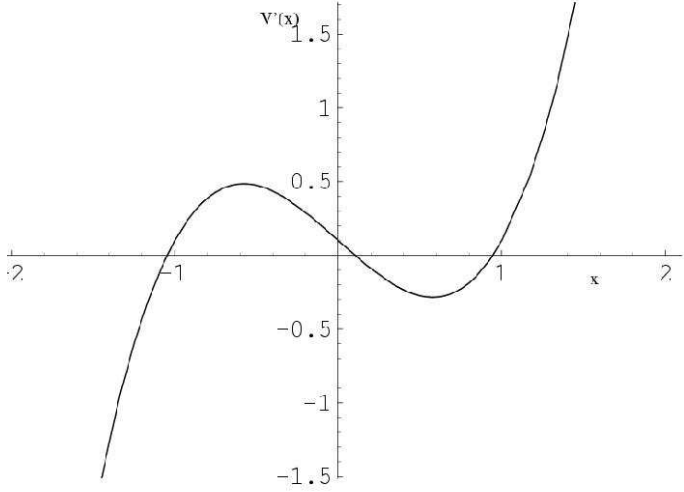}

\caption{ Derivative of the Asymmetric Potential 
$ \widetilde{V}^\prime (x) $ }
\label{fig3b} 

\end{figure}

(i). The saddle-point is solution of $ V^\prime (x) = a $ see 
     Fig. \ref{fig3b}.

(ii). If $a$ is positive we have three solutions but the action is lowest
      at $x_3$.

(iii). $x_3$ is still the leading saddle-point solution for $a<0$.

Therefore the behaviour of $m_n (a)$ for small $a$ is independent of 
sign $a$. This corresponds to the case studied in ref. \cite{aa96} where
the difference in the depth of the asymmetric wells is large.

(II). However if $V$ is symmetric, 
example: $V(x)={-1 \over 2}x^2 + {g \over 4}x^4$, when 
$a\rightarrow 0$ the saddle-points are  
\beq
x_c = \pm {1 \over \sqrt{g}}+{a\over 2}+O(a^2)
\eeq
($x\approx 0$ has a higher action) then
\beq
S(x_c) = {1 \over 2g} \mp {a \over \sqrt{g}}.
\eeq

The integral $m_n$ is dominated by

\beql
x &=& + { 1 \over {\sqrt{g}} }+{a \over 2} ~ for ~ a > 0 \nonumber \\
&=& - { 1 \over {\sqrt{g}} }+{a \over 2} ~ for ~ a < 0.
\eeql 

The moments are thus given by

\beql
m_n &=& {1 \over {g^{n\over 2}}} e^{-N\over {2g}} 
e^{+aN \over {\sqrt{g}}}\sqrt{2\pi\over 3N} ~ for ~ a > 0 \nonumber \\
&=& ({-1 \over {\sqrt{g}}})^n e^{-N\over {2g}} 
e^{-aN \over {\sqrt{g}}}\sqrt{2\pi\over 3N} ~ for ~ a < 0.
\eeql

For $n$ even the two results are the same; but for $n$ odd we get opposite
signs. Note that the $Z_2$ symmetry would say that $m_n=0$ for n odd and 
$a\rightarrow 0$. The set of moments 
would be $2^{N\over 2}$ corresponding to the number
of different possible moments (only the odd moments are different for 
different $n$). 

(III). We have to check whether the non-uniformity of the limits 
$N\rightarrow \infty$, $a\rightarrow 0$ may be present if $V$ is
non-symmetric but has two wells of equal depths.

\begin{figure} 

\leavevmode
\epsfxsize=4in
\epsffile{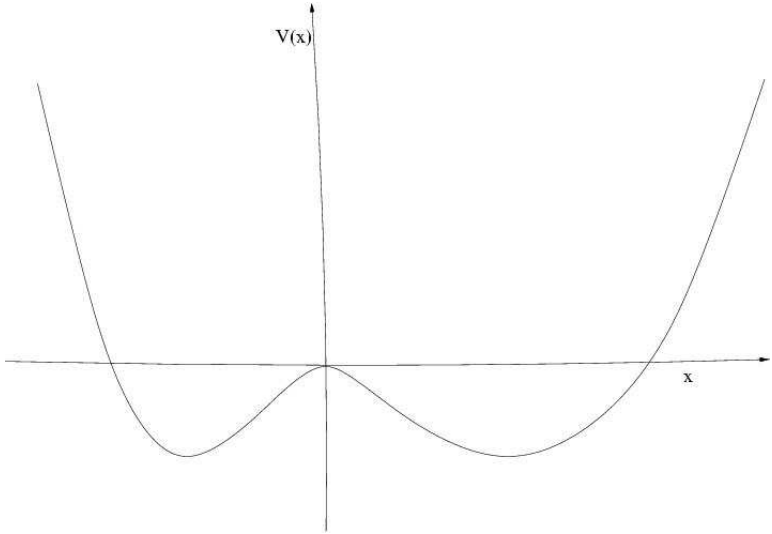}

\caption{ The Asymmetric Potential $V(x)$ 
With Two Wells Of Equal Depths }
\label{fig4a} 

\end{figure} 

The same series of arguments follow through for the asymmetric potential 
with two-wells of equal depths as for the purely symmetric potential. Hence
there would be multiple solutions of the same multiplicity $2^{N\over 2}$
in the moments for this problem as well. This is the situation considered 
in ref. \cite{bde00},
(though here only one of the $2^{N\over 2}$, the symmetric solution, as     
is refered to in ref. \cite{bdjt93} was considered) and arrive at the same 
symmetric answer as ref. \cite{bde00} where they make the unequal wells equal 
(asymmetry tending to zero limit).        

\section{First Steps In Counting Multiple Solutions}
\label{count}

Let us reformulate the problem in a slightly different way to
enable counting and bring out some novel results in a 
form easily comparable to formulae in ref. \cite{z97}. 
Consider the measure

\beq
Z^{-1} exp (-N tr V(M) + N tr MA) d^{N^2}M
\eeq

here $V$ is an arbitrary polynomial, and $A=diag(a_0,....,a_{N-1})$
can be assumed diagonal.

One diagonalizes M: if $M=\Omega \Lambda \Omega^{\dagger}$ where
$\Lambda = diag(\lambda_0,....,\lambda_{N-1})$, the integral over
$\Omega$ is the usual Itzykson-Zuber integral on the unitary group
and one finds:
 
\beq
\rho_N (\lambda_0,\lambda_1,...,\lambda_{N-1}) =
Z^{-1} \Delta (\lambda_i) 
{{det(exp N\lambda_j a_l})\over {\Delta (a_l)}}
exp \left( -N \sum_{i=0}^{N-1} V(\lambda_i) \right).
\eeq

Replacing powers of $\lambda$ in the Van der Monde with the orthogonal
polynomials $P_k (\lambda)$ of the measure $exp(-NV(\lambda)) d\lambda$.
The partition function $Z$ can then be expressed as:

\beql
Z &=& {N!\over \Delta (a_l)}\int \prod_{i=0}^{N-1} d\lambda_i 
det(P_k(\lambda_i)) 
exp N\sum_{i=0}^{N-1} (-V(\lambda_i)+a_i\lambda_i)\nonumber\\
&=& {N!\over \Delta (a_l)} det \left ( \int d\lambda P_k (\lambda)
exp N (-V(\lambda)+a_l\lambda) \right )
\label{partition}
\eeql

$\rho_N$ then becomes

\beql
\rho_N (\lambda_0,\lambda_1,...,\lambda_{N-1}) &=&
{1\over N!} {{det(P_k(\lambda_i))_{i,k=0...N-1} 
det(exp N a_l\lambda_j)_{j,l=0...N-1}}\over
{det(\int d\lambda P_k(\lambda) 
exp N(-V(\lambda)+a_l\lambda))_{k,l=0...N-1}}}\nonumber \\
exp \left( -N \sum_{i=0}^{N-1} V(\lambda_i) \right).
\label{rhoN1}
\eeql

This formula has a simple structure. On introducing the 
functions $F_k (\lambda) = h_k^{(-{1\over 2})} P_k (\lambda) 
exp {(-{N\over 2}V(\lambda))}$ and $G_l(\lambda)= 
exp\left ( Na_l\lambda-{N\over 2}V(\lambda) \right )$ we have

\beql
\rho_N (\lambda_0,\lambda_1,...,\lambda_{N-1}) 
={1\over N!}{ {det(F_k(\lambda_i))_{i,k=0...N-1}
det(G_l(\lambda_j)_{j,l=0...N-1}} \over 
{det( \int d\lambda F_k(\lambda) G_l (\lambda) )_{k,l=0...N-1}} }.
\label{rhoN2}
\eeql

The matrix $(\int d\lambda G_l (\lambda) F_k (\lambda))_{l,k=0...N-1}$
has an inverse $\alpha_{kl}$. Putting the three determinants together
we get:

\beq
\rho_N (\lambda_0,\lambda_1,...,\lambda_{N-1}) 
= {1\over {N!}} det (K(\lambda_i,\lambda_j))_{i,j=0...N-1}
\eeq

where 

\beq
K(\lambda,\mu)=\sum_{k,l=0}^{N-1} 
F_k (\lambda) \alpha_{kl} G_l (\mu).
\label{kernel}
\eeq 

The kernel satisfies the property:

\beq
[K*K](\lambda,\rho)=K(\lambda,\rho).
\eeq

Thus we obtain the determinant formulae

\beq
\rho_n (\lambda_0,\lambda_1,....,\lambda_{n-1}) 
= {(N-n)!\over N!} det (K(\lambda_i,\lambda_j))_{i,j=0...n-1}
\eeq

for any $n\leq N$. The kernel $K$ has the form 

\beq
K(\lambda,\mu) = \sum_{k=0}^{N-1} F_k (\lambda) \hat{F}_k (\mu)
\label{kernel}
\eeq

with $\hat{F}_k (\mu)= \sum_l \alpha_{kl} G_l (\mu)$; but 
$\hat{F}_k\neq F_k$. Thus $K$ is not symmetric. In order to
get further properties for $K$ we consider the integral

\beql
I &=& \int d\lambda \left ( G_l (\lambda) 
F_k (\lambda) \right )_{l,k=0...N-1} \nonumber \\
&=& \int d\lambda {P_k (\lambda)\over \sqrt{h_k}}
exp (N(-V(\lambda)+a_l\lambda))\nonumber \\
&=& {1\over \sqrt{h_k}} \int d\lambda  \sum_{i=0}^k C_i \lambda^i
exp (N(-V(\lambda)+a_l\lambda))\nonumber \\
&=& {1\over \sqrt{h_k}}  \sum_{i=0}^k C_i  
\int d\lambda \lambda^i exp (N(-V(\lambda)+a_l\lambda))\nonumber \\
&=& {1\over \sqrt{h_k}}  \sum_{i=0}^k C_i m_i 
\eeql
\noindent
$m_i$ are the moments. For symmetric potential $V(\lambda)$ the above
expression becomes ( using the expression for the moments found in the
previous section )

\beql
I &=& \int d\lambda \left ( G_l (\lambda) F_k (\lambda)\right )_{l,k=0...N-1} 
\nonumber \\
&=& {1\over \sqrt{h_k}} \sum_{i=0}^k C_i g^{i\over 2} 
e^{-{N\over 2g}+{ a_l N\over \sqrt{g} }} \sqrt{2\pi\over 3N} ~~~ a_l > 0
\nonumber \\
&=& \alpha_{kl}^{-1} \nonumber \\
&=& {1\over \sqrt{h_k}}  \sum_{i=0}^k C_i (-{1\over \sqrt{g}})^i 
e^{-{N\over 2g}-{ a_l N\over \sqrt{g} }} \sqrt{2\pi\over 3N} ~~~ a_l < 0
\nonumber \\
&=& \alpha^{\prime -1}_{kl}. 
\label{glfk}
\eeql

Summarizing

\beql
I = \left\{ \begin{array}{ll}
   \alpha^{-1}_{kl}, & a_l > 0 \nonumber \\
      
    \alpha^{\prime -1}_{kl}, & a_l < 0 
    
    \end{array}
\right.
\eeql

Recall that $x_c = \pm {1 \over \sqrt{g}}+{a\over 2}$ thus only for
$\pm {1 \over \sqrt{g}} \ge {a\over 2}$ the above result holds i.e.
the integral eq. (\ref{glfk}) has two values depending on whether $a_l>0$
or $a_l<0$. Whereas for $\pm {1 \over \sqrt{g}} \le {a\over 2}$ the usual 
single well result as given in ref. \cite{z97} is found.   

From the equation for $K (\lambda,\mu)$ i.e. eq. (\ref{kernel})
which depend on the integral eq. (\ref{glfk}) through a sum
it may be possible that there are $2^N$ solutions for certain kernels
this would corresponds to an exponentially large number of solutions 
depending on the path or different combinations of $a_l$ taken. 
For $\rho_N (\lambda_0,...,\lambda_{N-1})$ 
and $Z$ i.e. eq.(\ref{partition})and eq.(\ref{rhoN1}) which are related
to $I$ through a determinant it is risky to consider the large
N behavior of $I$ before computing $det_{[N\times N]} I$. Counting
at the level of $K (\lambda,\mu)$, $\rho_N (\lambda_0,...,\lambda_{N-1})$, $Z$
and the free energy still remains an open one and needs a non-perturbative
treatment (as shown in ref. \cite{bdjt93}). This will be pursued in
a future work. 

\section{An Explicit Calculation Of The Integral Eq. (\ref{glfk})
For The Double-Well Problem}

For the double-well matrix model the orthogonal polynomials are
not known polynomials but we do know the form for the polynomial
at large $N$ ie when $ (N-n) \approx O(1) $. The polynomials are
given by  
 
\beq
\psi_n (\lambda) = {1\over \sqrt{f}} 
\left [ \cos ( N \zeta - (N-n)\phi + \chi + (-1)^n \eta ) (\lambda)
+ O ({1\over N}) 
\right ]
\eeq

where $f, \zeta, \phi, \chi$ and $\eta$ are functions of $\lambda$ and
are given by

\beql
f(\lambda) &=& {\pi\over 2 \lambda} {(b^2-a^2)\over 2} 
\sin 2\phi (\lambda) \nonumber \\
\zeta^{\prime} (\lambda) &=& - \pi \rho (\lambda) \nonumber \\
\cos 2 \phi (\lambda) &=& {\lambda^2-{(a^2+b^2)\over 2}\over 
{(b^2-a^2)\over 2}} \nonumber \\
\cos 2 \eta (\lambda) &=& b {\cos \phi (\lambda) \over \lambda} 
\nonumber \\
\sin 2 \eta (\lambda) &=& a {\sin \phi (\lambda) \over \lambda}
\nonumber \\
\chi (\lambda) &=& {1\over 2} \phi (\lambda) - {\pi\over 4}
\label{fzpec}
\eeql

Let us consider the eq. (\ref{glfk}) with the above asymptotic
ansatz for $\phi_k$ for large $k$ then

\beql
I = \int P_k (\lambda) e^{-N (V(\lambda) - a\lambda)} \nonumber \\
P_k (\lambda) e^{-{N\over 2} V(\lambda)} = \sqrt{h_k} 
\psi_k (\lambda) \nonumber \\
I = \sqrt{h_k} Re \int {d\lambda \over \sqrt{f(\lambda)}} 
e^{i(N\zeta - (N-k)\phi + \chi + (-1)^k \eta)} e^{-N({1\over 2}{1\over 2}
\mu \lambda^2 -a \lambda)} \nonumber \\
= \sqrt{h_k} Re \int {d\lambda} e^{N ({{1\over 2N} \ln f(\lambda) + i \zeta 
+ i\gamma_{N,k}{\phi\over N} + i (-1)^k {\eta\over N} 
- {1\over 4} \mu \lambda^2 + a\lambda - {\pi \over 4N} ) }} 
\eeql

Where $\gamma_{N,k}$ is given by $-(N-k)+{1\over 2}$.  
In the saddle point approximation the exponent $S(\lambda)$ is
to be minimized. The action $S(\lambda)$

\beq
S(\lambda)={i\gamma_{N,k} \phi (\lambda) \over N} 
+ i \zeta 
+i (-1)^k {\eta (\lambda)\over N} - {1\over 4} \mu \lambda^2 + a \lambda + 
{1\over 2N} \ln f(\lambda)  
\eeq

will have a first derivative which vanishes as shown below

\beql
{i\gamma_{N,k} \phi^{\prime} (\lambda) \over N} + i \zeta^{\prime}
+ i (-1)^k {\eta^{\prime} (\lambda) \over N} 
- {1\over 2} \mu \lambda + a + {1\over {2N f(\lambda)}} f^{\prime} (\lambda)
= S^{\prime} (\lambda) = 0 \nonumber \\ 
{i\gamma_{N,k} \phi^{\prime} (\lambda) \over N} - i \pi \rho (\lambda) 
+ i (-1)^k {\eta^{\prime} (\lambda) \over N}
- {1\over 2} \mu \lambda + a + {1\over {2N f(\lambda)}} f^{\prime} (\lambda)
= 0.\nonumber \\ 
\eeql

Where we have used the relation for $\zeta$ in terms of $\rho$ from 
eq. (\ref{fzpec}). Solving for the density $\rho (\lambda)$ we get

\beql
\rho (\lambda) = {i\over \pi} ({1\over 2} \mu \lambda - a ) + {{\gamma_{N,k}
\phi^{\prime} (\lambda)} \over {\pi N}} - {i\over {2N \pi f(\lambda)}} 
f^{\prime} (\lambda) + (-1)^k {\eta^{\prime} (\lambda) \over {\pi N}}
\eeql

For the symmetric potential $\rho (\lambda) = {1\over \pi} 
\sqrt{\lambda^2-b^2}$, in the large N limit neglecting the last 
terms, as the equation for $\lambda$ is quadratic there are two solutions
to the equation as shown below

\beql
\lambda^2 &-& { {\mu \lambda a}\over {({1\over 4}\mu^2+1)} } 
+ { {a^2-b^2}\over { ({1\over 4}\mu^2+1) } } = 0 \nonumber \\
\lambda_{\pm} &=& { {\mu a }\over {2({1\over 4}\mu^2+1)} } \pm
{1\over 2} \sqrt{\left ( {{\mu a}\over {{1\over 4}\mu^2+1}} \right )^2 
- { {4 (a^2-b^2)} \over {({1\over 4}\mu^2+1)}} }.
\eeql

Thus in the saddle point approximation the integral $I$ for large $k$
becomes

\beql
I_{\pm} &=& I_0 P_k (\lambda_{\pm}) 
e^{-N ( V(\lambda_{\pm})-a\lambda_{\pm})} + h.o. \nonumber \\
\eeql

Where $I_0$ is a constant.
Hence we have shown in an explicit example for the symmetric double well
potential that the integral eq.(\ref{glfk})
for large $k$ in the saddle point approximation has two solutions, 
which solution is choosen depends on whether $a \ge or \le 0$. 
This result indicates the possibility that the kernel, partition function, 
free energy can have $2^N$ solutions depending on the path $\{ a_l \}$ 
taken as these functions all depend on the integral $I$, 
eq.(\ref{glfk}). Thus here evidence is presented that there exists an 
exponentially large number of solutions, i.e. $e^{N\ln 2}$, 
in the double well matrix 
models depending upon the path taken in parameter space $\{ a_l \}$.
It will be interesting to explore the possibility that these exponentially
large number of solutions correspond to the metastable solutions of the
supercooled p-spin glass that these random matrix models map into.

\section{Conclusions}

\label{conclusions}

We have been able to map the problem of counting the number of multiple
solutions found in ref. \cite{bdjt93} to a moment problem. The multiple 
solutions
were discovered in the recurrence coefficients of the orthogonal polynomials
in ref. \cite{bdjt93}. It was known that there are an infinite number of 
solutions.
The counting problem is mapped onto counting the number of ways to get 
different moments. The set of moments grows exponentially as 
$2^{N\over 2}$. In order to show this we have to introduce a small
perturbation which breaks $Z_2$ symmetry into the moment integral and then
take the small asymmetry parameter to zero ( which we call tapping the
matrix ). As an added bonus we are able to understand some
of the puzzles and controversies that are found in ref. \cite{bdjt93} 
and studied in ref. \cite{d97,bd99}. The counting at the level of
the kernel, $\rho_N (\lambda_0,...,\lambda_{N-1})$, $Z$
and the free energy still remains an open one and needs a non-perturbative
treatment. This will be pursued in a future work. 

The number of moments in these random matrix models are 
exponentially rising with $N$. These matrix models are connected 
with the high temperature phase of structural glasses as has been 
discussed in refs. \cite{d02,ckpr95}. There could be interesting 
properties of the 
supercooled liquid phase which may be explored analytically in these
simple models. For example it will be worthwhile to study how the metastable
states of the liquid are related to the different paths of taking the 
small perturbation parameter, as introduced here, to zero. Future work
on finding barrier heights is underway. 
 
\section{Acknowledgements}
\label{acknowledgements}

Jorge Kurchan and Letitia Cugliandolo are 
gratefully acknowledged for the communication of the 
ref. \cite{sw00} and for introducing ND to the interesting subject of 
spin glasses and granular materials along with the dynamical and
Edwards temperature. 
  


\begin{thebibliography}{999}


\newcommand{\NP}[3]{{\it Nucl. Phys. }{\bf B#1} (#2) #3}
\newcommand{\PL}[3]{{\it Phys. Lett. }{\bf B#1} (#2) #3}
\newcommand{\PR}[3]{{\it Phys. Rev. }{\bf #1} (#2) #3}
\newcommand{\PRL}[3]{{\it Phys. Rev. Lett. }{\bf #1} (#2) #3}
\newcommand{\IMP}[3]{{\it Int. J. Mod. Phys }{\bf #1} (#2) #3}
\newcommand{\MPL}[3]{{\it Mod. Phys. Lett. }{\bf #1} (#2) #3}
\newcommand{\JP}[3]{{\it J. Phys. }{\bf A#1} (#2) #3}


\bibitem{d02} N. Deo, \PR {E 65}{2002}{056115}.

\bibitem{bd99} E. Brezin and N. Deo,
\PR{E 59}{1999}{3901}.

\bibitem{sw00} M. Soljacic and F. Wilczek, \PRL {84}{2000}{2285}.

\bibitem{mg91} M. L. Mehta {\it{Random matrices}} (Academic Press, 1991);
T. Guhr, A. Mueller-Groeling and H. A. Weidenmueller, Phys. Rep. {\bf 299}, 
(1998), 189.

\bibitem{oz01} H. Orland and A. Zee, cond-mat/0106359.

\bibitem{bs99} L. Laloux, P. Cizeau, J. P. Bouchaud and M. Potters,
\PRL {83}{1999}{1467}; V. Plerou, P. Gopikrishnan, B. Rosenow, L. A. N.
Amaral and H. E. Stanley,\PRL {83}{1999}{1471}.

\bibitem{ckpr95} L. F. Cugliandolo, J. Kurchan, G. Parisi, and F. Ritort,
\PRL{74}{1995}{1012};
G. Parisi, Statistical Properties of Random Matrices
and the Replica method, cond-mat/9701032.

\bibitem{dl01} D. S. Dean and A. Lefevre, \PRL {86 (25)}{2001}{5639}.

\bibitem{bdjt93} R. C. Brower, N. Deo, S. Jain and C. I. Tan,
\NP{405}{1993}{166}.

\bibitem{aa96} G. Akemann and J. Ambjorn, \JP{29}{1996}{L555};
G. Akemann, \NP{482}{1996}{403} and \NP{507}{1997}{475}.

\bibitem{bde00} G. Bonnet, F. David and B. Eynard, \JP {33}{2000}{6739}.

\bibitem{z97} P. Zinn-Justin, \NP {497}{1997}{725}.

\bibitem{d97} N. Deo, \NP{504}{1997}{609}.





\end{thebibliography}
\end{document}